\documentclass[aps,prl,showpacs,,twocolumn,superscriptaddress]{revtex4}

\usepackage{graphicx}
\usepackage{amsmath}
\begin{document}

\title{Manifestations of electron interactions in the photogalvanic effect in chiral nanotubes}

\author{Raphael Matthews}

\affiliation{The Racah Institute of Physics, The Hebrew University,
Jerusalem 91904, Israel}

\author{Oded Agam}
\affiliation{The Racah Institute of Physics, The Hebrew University,
Jerusalem 91904, Israel}

\author{Anton Andreev}
\affiliation{Department of Physics, University of Washington, Seattle, Washington 98195, USA}

\author{Boris Spivak}
\affiliation{Department of Physics, University of Washington, Seattle, Washington 98195, USA}

\date{\today}

\begin{abstract}
Carbon nanotubes provide one of the most accessible experimental realizations of one dimensional electron systems. In the experimentally relevant regime of low doping the Luttinger liquid formed by electrons may be approximated by a Wigner crystal. The crystal-like electronic order suggests that nanotubes exhibit effects similar to the M\"ossbauer effect where the momentum of an emitted photon is absorbed by the whole crystal.  We show that the circular photovoltaic effect in chiral nanotubes is of the same nature. We obtain the frequency dependence of the photovoltage and characterize its singularities in a broad frequency range where the electron correlations are essential.  Our predictions provide a basis for using the photogalvanic effect as a new experimental probe of electron correlations in nanotubes.
\end{abstract}

\pacs{78.67.Ch}
\maketitle

\section{I. Introduction}

Carbon nanotubes (CN)  can be viewed as long cylinders made from a graphene sheet \cite{Ijima,Phys,Review}. Depending on the way this sheet is rolled up, the cylinder may have chiral (i.e., helical) structure. Chirality implies the existence of the circular photogalvanic effect (CPGE), where a circularly polarized electromagnetic wave induces a dc current. The magnitude of the photo-induced voltage is finite even when the momentum of the photon is negligible. This effect was predicted in Refs.~\cite{IvchenkoPikus,belinicher} and since then was investigated both experimentally and theoretically - see, for example, Refs.~\cite{SturmanFradkin,IvchPik,Ganichev,GanichevDanilov} and references therein.

Usually the circular photogalvanic effect in non-centrosymmetric media arises due to a spin-orbit interaction. In carbon nanotubes the spin-orbit interaction is weak and the circular photogalvanic effect arises because the electron motion along the tube axis and circumference is strongly mixed in chiral nanotubes. The single particle theory of chirality related optical effects in carbon nanotubes has been developed in Ref.~\cite{IS}. This theory holds for non-degenerate electrons in semiconductor nanotubes or for metallic ones at high enough frequencies of the electromagnetic field.

However, at low temperatures the single particle description of one-dimensional interacting electron systems, such as metallic CN, is invalid. Being a quasi-one-dimensional system, CN  exhibits distinctive features which result from the strong correlations of the electrons in the system \cite{Gi}, for instance, the Fermi edge singularity which is manifested by a characteristic power-law singularity of the tunneling density of states. The photovoltaic effect in this regime was not investigated, and will be studied here.

The paper is organized as follows. In Sec. II we review the non-interacting theory of the CPGE in CN. In Sec. III we study the problem with electron-electron interactions.  Next, in Sec. IV,  we discuss the limit where interactions are strong enough so that the electronic correlations are governed by Wigner crystal order. We draw parallels between the CPGE and the M\"{o}ssbauer effect, and show that the photovoltage,
as  a function of the frequency of  the electromagnetic wave, exhibits a series of distinct singularities.
Our results are summarized in Section V.
\begin{figure}
\centerline{ \resizebox{0.47 \textwidth}{!}
{\includegraphics{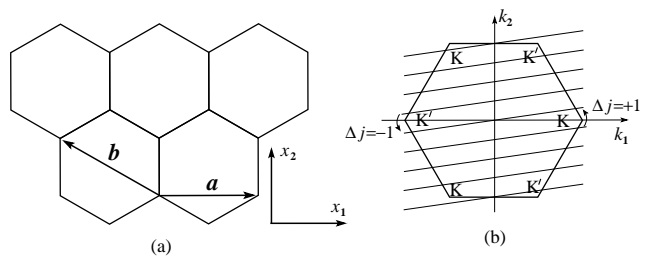}}} \caption[] {(a) The honeycomb lattice of graphene and its primitive lattice vectors $\bf{a}$ and $\bf{ b}$.  (b) The Brillouin zone of graphene. $K$ and $K'$ denote the Dirac points of the energy spectrum. The straight lines cut the zone at the discrete set of allowed transverse momenta. This set is determined by the way a graphene sheet is rolled into a tube, such that the wave function is invariant with respect to the translation vector ${\bf L}= n{\bf a}+ m {\bf b}$. The example shown here corresponds to $(n,m)=(5,12)$. The lowest energy excitation near the $K$ and $K'$ points of the Brillouin zone are associated with a change of the angular momentum by $\pm \hbar$, thus a circularly polarized light directed along the tube excites electrons only near one of the Dirac points.}
\end{figure}

\section{II. Single particle picture}

Let us begin with a brief review of the noninteracting theory of the CPGE in CN's. Fig.~1(a) presents the honeycomb lattice structure of a graphene sheet, and its two primitive lattice vectors ${\bf a}=a( 0,1)$ and ${\bf b}=a(-1,\sqrt{3})/2$, where $a=2.49$\AA  ~is the lattice constant. A single-walled CN rolled up from such a sheet is defined by two integers $n$ and $m$ which specify the translation vector ${\bf L}= n{\bf a}+ m {\bf b}$ that wraps around the cylinder. The requirement that  the wave function be invariant with respect to  a translation by ${\bf L}$,  $\psi({\bf x}+ {\bf L}) = \psi({\bf x})$, imposes  a quantization condition on the component of the wave vector in the transverse direction $k_\perp |{\bf L}| =2 \pi j$, where $j$ is an integer.  Fig.~1(b) shows the Brillouin zone of graphene. The corners of this zone are Dirac points where the spectrum is degenerate and the two inequivalent points are denoted by $K$ and $K'$. The straight lines cutting through the zone represent the aforementioned quantization condition. Each line, i.e, each value of $j$, defines a subband of the nanotube spectrum which corresponds to a definite angular momentum of the electron around the cylinder (see, for example, Refs.~\cite{Tasaki,Ajiki,IS}).

A characteristic feature of chiral nanotubes is that the minima of subbands with different values of angular momentum are shifted with respect to each other by a momentum $\delta p$, as illustrated in Fig.~2. To be concrete we consider the situation where the lowest conduction subband with angular momentum $\hbar\, j$ is partially occupied (say due to  doping by an an external gate), and the angular momentum of the first unoccupied band of the $K$ valley is $\hbar(j+1)$.  Now let us assume that a circularly polarized electromagnetic wave is radiated in the direction of the tube. Absorption of a photon will be accompanied by a unit change of the angular momentum, which at low enough frequencies might take place only in one valley, say K, since the corresponding transition in the opposite (K') valley is forbidden by conservation of angular momentum; see Fig.~1(b). A finite value of the photovoltage emerges from the asymmetry of velocities and light absorption probabilities for right- and left- moving electrons in the $K$ valley. The single particle absorption threshold $\omega_{c}^{sp}$  is determined by the electron band structure and is illustrated in Fig.~2. Because of the constant density of states at the Fermi level, the photo-voltage $V(\omega)$ as a function of the frequency $\omega$ experiences a jump from zero to a finite value at the threshold frequency $\omega=\omega_{c}^{sp}$.

\begin{figure}[ptb]
\includegraphics[width=7.0cm]{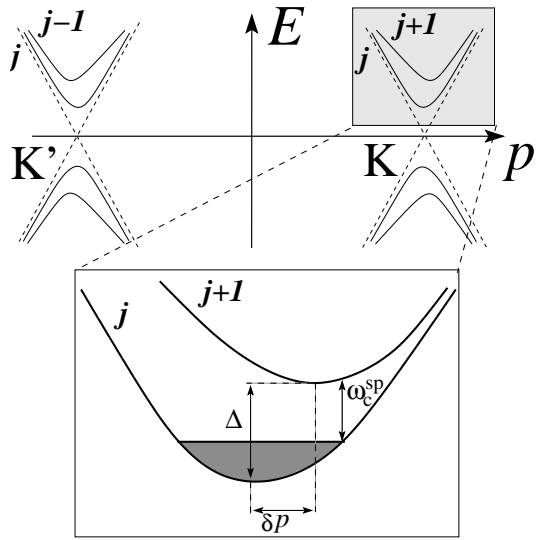}
 \caption{Subbands of chiral nanotubes demonstrating the asymmetric band structure near each valley. The shaded region in the lower box represents the occupied states within the single particle picture.}
\end{figure}

\section{III. Interacting particles picture}

Generally, the frequency $\omega_c$ of the  many-body absorption threshold is different from the single particle threshold $\omega_{c}^{sp}$. At sufficiently large values of $\delta p$ the absorption threshold can be significantly lower. It corresponds to an indirect transition in which one electron is transferred to a state near the bottom of the upper subband. The total momentum of the system does not change as a result of photoabsorption. Therefore a momentum $ - \delta p$ is imparted to the electrons in the lower (partially occupied) subband. Since the velocity of the excited electron in the upper band is small, only the electrons in the lower band contribute to the photocurrent. The photovoltage $V(\omega)$ necessary to nullify the current is determined by the momentum flux required to compensate for the flux of momentum to the lower band electrons due to photon absorption. Denoting the photon absorption rate by  $W(\omega)$, we obtain for the photo-induced voltage
\begin{equation}\label{voltage}
V(\omega)=\frac{\delta p}{n e}\,  W(\omega),
\end{equation}
where $n$ is the electron density in the lower band, and $e$ is the electron charge.
The precise criteria of validity of Eq.~(\ref{voltage})
depend on the value of $r_{s}$ and will be discussed below.  Here $r_{s}=U/E_{kin}$ is the ratio between the characteristic potential, $U\sim e^2n$, and  the kinetic, $E_{kin} \sim(\hbar n)^{2}/m$ ($m$ being the electron mass) energies of the electrons.

Another manifestation of the many body effects is that near the threshold, $|\omega -\omega_{c}|\ll \omega_{c}$, the photovoltage exhibits a power law singularity $V(\omega)\sim (\omega-\omega_{c})^{\eta_0}$ characteristic of one dimensional (1D) systems. We will show that $\eta_0>0$, for $\delta p \gg \hbar n r_s^{1/4}$, and  $\eta_0 <0$  when  $\delta p \ll \hbar n r_s^{1/4}$. Thus interactions suppress the photogalvanic effect near the threshold in the first case and enhance it in the second. We shall also discuss additional singularities in $W(\omega)$ which occur above the threshold at intervals of the Debye frequency of the Wigner crystal and are associated with the generation of hard plasmons.

In principle, recombination processes create an additional contribution to the electron distribution function in the lower band which is asymmetric in $p$ and  which affects the photovoltage $V(\omega)$. In this case however,  the singularity of  $V(\omega)$ at the absorption threshold remains intact. Recombination with a large energy transfer usually takes place through a multiphonon emission via deep impurities which are associated with a large momentum transfer to individual phonons. Therefore the momentum asymmetry of the distribution function in the lower band associated with such processes is small and can be neglected.

The significance of many-body effects for the absorption and the photovoltaic effect can be appreciated even in the framework of perturbation theory using simple kinematic considerations. Let us assume that $\delta p$ is larger than the characteristic momenta of electrons in the lower band. Consider an absorption process in which one electron is transferred to a state near the bottom of the upper band and $N$ electrons in the lower band receive a recoil momentum of order $ p^* \approx - \delta p/N$ and energy $\epsilon^* \sim (p^*)^2/2m$. The energy of such a photoexcitation is $\hbar \omega_N \approx \Delta + (\delta p)^2/(2m N)$, where $\Delta$ is the energy difference between the bottoms of the bands (Fig.~1). It decreases with increasing $N$ and may be significantly lower than the single particle threshold $\omega_{c}^{sp}\approx \omega_{N=1}$. If  $\epsilon^*\gg  \max[U,E_{F}]$, regardless of the  details of the electron wave function in the ground state, the excited electrons may be treated as free particles and the absorption rate $W(\omega)$ can be calculated using the $N$-th order of the perturbation theory in a way similar to that of Refs.~\cite{Voloshin92,Brown}. The increase of $N$ with decreasing frequency can be considered as a precursor of the Luttinger liquid regime. Practically, however, the condition $(\delta p)^{2}/( 2m) \gg \max[U,E_{F}]$ is difficult to realize in carbon nanotubes, and therefore there is no parametrically big interval of frequencies where the perturbation theory works. For this reason, below, we focus our attention on the case where frequency is close to the threshold.

\section{IV. The Wigner crystal limit}

 Consider the experimentally relevant regime of low electronic densities $n$, where the ratio of the interaction energy to the kinetic energy  of the electrons is large, $r_s \gg 1$. In this case exchange processes associated with the tunneling of electrons are exponentially suppressed in $r_s$,  and will therefore be neglected in what follows. In this approximation electrons may be labeled by the site number $\nu$ in the Wigner crystal lattice.

 The wave functions of electrons localized near a given site are superpositions of Bloch functions either in the valley $K$ or $K'$. The two possibilities are realized with equal probability. In a broad frequency range electrons excited by light into the upper subband remain localized at their original sites of the Wigner crystal. For a circularly polarized light only electrons from  valley $K$ can participate in the absorption. Thus the wave functions of excited electrons are superpositions of Bloch wave functions of the upper band in the $K$ valley. Since they are localized near the Wigner crystal sites, their average momentum must be $\delta p$; see Fig.~1. Conservation of momentum dictates that the opposite momentum $-\delta p$ must be transferred into the collective motion of the Wigner crystal, which leads to Eq.~(\ref{voltage}). A similar transfer of the recoil momentum from a given atom to the whole crystal occurs during the emission of $\gamma$ rays. One of the striking manifestations of the transfer of the recoil momentum to the collective motion in three dimensional crystals is the M\"ossbauer effect~\cite{Lipkin} in which no phonons are excited during a $\gamma$-ray emission. The probability for such a transition is given by the Debye-Waller factor $\exp[- (\delta p)^2 \langle x^2 \rangle/\hbar^2]$, where $\delta p$ is the recoil momentum and  $\langle x^2 \rangle$ is the variance of the displacement of the nucleus from its equilibrium position. In a one dimensional system the latter diverges logarithmically with the system size and the M\"ossbauer effect is impossible. However light absorption with the emission of collective excitations (plasmons) remains possible. The probability of plasmon emission determines the absorption rate above the threshold, $\omega> \omega_c$.
Though these plasmons carry the excess energy, their average momentum is zero. Thus each photon absorption provides a momentum $\delta p$ which is transferred to the Wigner crystal of the electrons in the occupied subband. Since the average velocity of the excited electron in the upper band is zero, we again arrive at the expression for $V(\omega)$ given by Eq.~(\ref{voltage}).

Near the absorption threshold the probability of interband transitions exhibits a power law singularity, which can be described phenomenologically in terms of a motion of a mobile impurity in a Luttinger liquid~\cite{Ogawa92,Balents00,Zvonarev07,Imambekov09,kamenevGlazman09,Pereira09,Mishchenko11}.
The strength of electron-electron interactions in the Wigner crystal regime enables us to determine the light absorption probability not only near the absorption threshold but in a much wider frequency range.

Although many of our conclusions have general character, we will consider the case $(\delta p)^2/2m \ll ne^2$ where the Wigner crystal may be viewed as a weakly anharmonic chain~\cite{Matveev2010}. We model our system by the Hamiltonian $H=H_0+ H_{ep}$, where the unperturbed Hamiltonian is
\begin{eqnarray}\label{eq:H_WC}
    H_0 = \sum_\nu
    \left(
               \begin{array}{cc}

                 \Delta+\frac{ (p_\nu- \delta p)^2}{2m} &  0 \\
                 0 & \frac{p_\nu^2}{2m}
               \end{array}\right)
                + \sum_{\mu >\nu}V(x_{\mu}-x_{\nu}).
\end{eqnarray}
Here $p_\nu$ and $x_\nu$ denote the momentum and position of the $\nu$-th electron, $\Delta$ is the energy shift between the bands,  and  $V(x)$ is the interaction potential. For simplicity the electron mass is assumed to be the same in both subbands. The matrix in Eq.~(\ref{eq:H_WC}) acts on the subband index. In the rotating wave approximation the interaction with the electromagnetic field is described by
\begin{eqnarray}\label{eq:H_ep}
    H_{ep} = D \sum_\nu (\sigma_\nu^+ a e^{-i\omega t} + h.c.).
\end{eqnarray}
Here $a$ is the photon annihilation operator, $\sigma_\nu^+$ is the raising operator of the $\nu$-th electron from the lower to the upper subband,  and  $D$ is the dipole matrix element of the transition.
The electron positions and momenta in a Wigner crystal regime can be expressed in terms of the plasmon annihilation
and creation operators, $b_q$ and $b_q^\dagger$,
\begin{subequations}
\begin{eqnarray}
x_\nu &=& \nu/n +\sum_q \sqrt{\frac{\hbar}{2mN\omega_q}}\, (b_q+b_{-q}^\dagger)
  e^{iq \nu}, \\
  p_\nu  &=& -i\sum_q \sqrt{\frac{\hbar m\omega_q}{2N}}\, (b_q-b_{-q}^\dagger)
  e^{iq \nu},
\end{eqnarray}
\end{subequations}
where $n$ and $N$ are the electron density and the number of sites, respectively, in the Wigner crystal, and $\omega _q$ is the frequency of the a phonon with (dimensionless)
wave number $q$ which stratifies the dispersion relation:
\begin{equation}
\omega_q^2 = \frac{2}{m} \sum_{\nu =1}^\infty  \left. \frac{ d^2 V(x)}{dx^2} \right|_{x= \nu /n} \left[ 1- \cos  (q\nu) \right].
\end{equation}
Denoting by $c$ and $c^\dagger$ the annihilation and creation of photoexcitations, respectively,  on site $\nu=0$, one can write the Hamiltonian in the form
\begin{eqnarray}\label{eq:H_WC_photoexcitation}
    H_0&=&\sum_q \hbar \omega_q \left( b^\dagger_q \!+\! \frac{ i\delta p ~ c^\dagger c  }{\sqrt{2m N \hbar \omega_q}}\right)\left( b_q \!-\! \frac{i \delta p~c^\dagger c  }{\sqrt{2m N \hbar \omega_q}}\right) \nonumber \\ &+& \Delta c^\dagger c,
\end{eqnarray}
where we have subtracted the zero point energy of the plasmons.
The absorption of photons happens independently at different lattice sites; thus, within the dipole approximation, it is given by the Kubo formula,
\begin{eqnarray} \label{eq:rate}
W(\omega) \!=\! \frac{N D^2}{\hbar^2} \mbox{Re}  \! \int_0^\infty dt e^{i \omega t} \langle [ c(t), c^\dagger (0)] \rangle,
\end{eqnarray}
where $\langle \cdots \rangle$  denotes thermal averaging.
To decouple the interaction between the plasmons and the excited particle we perform a unitary transformation of the operators
\begin{equation}\label{eq:unitary}
    \hat{U}=\exp\left(i\delta p~ c^\dagger c   \sum_q \frac{b_q+b_{-q}^\dagger}{\sqrt{2m N\hbar\omega_q}}\right).
\end{equation}
In the transformed basis the Hamiltonian takes the form
 \begin{equation}
 \hat{U}^\dagger H_0 \hat{U} = \sum_q \hbar \omega_a b^\dagger_q b_q+ \hbar \omega_c c^\dagger c,
 \end{equation}
 while the creation and annihilation operators are
  \begin{equation}
 \hat{U}^\dagger c \hat{U} =
 c e^{i\delta p x_0/\hbar}, ~~~\mbox{and}~~~ \hat{U}^\dagger c^\dagger \hat{U} = c e^{-i\delta p x_0/\hbar}.
 \end{equation}
 Thus in the limit of zero temperature Eq.~(\ref{eq:rate}) reduces to
\begin{eqnarray} \label{eq:W}
W(\omega) =  \frac{N D^2}{\hbar^2} \int_{-\infty}^\infty \!dt \exp \left(i (\omega\!-\!\Delta/\hbar) t - \frac{(\delta p)^2 F(t)}{2 m \hbar \omega_D} \right),
\end{eqnarray}
where $\omega_D= \omega_\pi$ is the Debye frequency, and
\begin{eqnarray}
F(t) =\int_0^{2 \pi} \frac{dq}{2\pi}  \frac{\omega_D }{\omega_q}  (1-e^{ -i \omega_q t}). \label{eq:F0}
\end{eqnarray}

\begin{figure}[ptb]
\includegraphics[width=7.9cm]{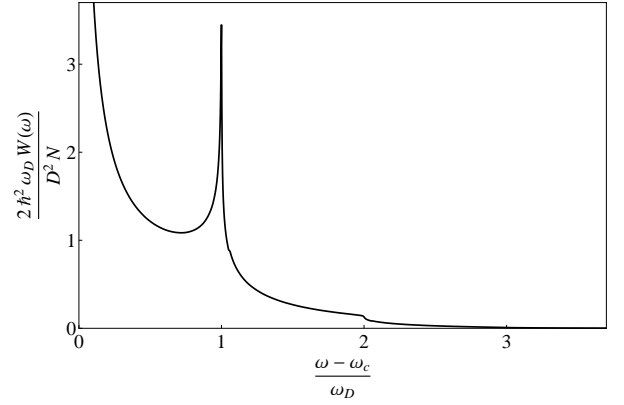}
 \caption{The absorption spectrum obtained by numerical integration of (\ref{eq:W}) for the case of short range interactions and a momentum shift $\delta p$, for which $\eta_0=-0.4$}
\end{figure}

The singular behavior of the absorption rate is determined by the long time asymptotic behavior of $F(t)$. In this limit the integral over $q$ may be evaluated by the stationary phase approximation. For a typical dispersion, the stationary phase is at $q=\pi$  and the dispersion, near this point, may be approximated as
\begin{equation}
\omega_q \simeq \omega_D+ \frac{1}{2}\omega''_\pi (q-\pi)^2,
\end{equation}
with $\omega''_\pi = d^2\omega_q/dq^2|_{q=\pi}$.  But there are also contributions from the end points of the integration interval, at $q=0$ and $q=2\pi$. Near these points, the  dispersion may be linearized. In particular, near $q=0$,
\begin{equation}
\omega_q \simeq \omega'_0 q,
\end{equation}
where $\omega'_0 = d\omega_q/dq|_{q=0}$, and similarly near $q=2\pi$,  $\omega_q \simeq \omega'_0 (2\pi-q)$. Evaluating the integral (\ref{eq:F0}) using these approximations we obtain
\begin{eqnarray}
F(t) \sim \frac{2\omega_D}{\pi \omega'_0} \ln (i \omega'_0 t)+ \gamma- \frac{ \exp( - i \omega_D t)}{\sqrt{ 2\pi i \omega''_\pi t}},
\end{eqnarray}
where $\gamma$ is a constant of order unity which depends on the precise form of $\omega_q$.  The first term of this asymptotic formula accounts for the soft plasmons generated by the excitation, while the last term is associated with the excitation of hard plasmons with energy near the plasmon Debye frequency $\omega_D$. Substituting this expression in (\ref{eq:W}) and expanding the hard plasmons contribution in a power series we find that the absorption exhibits singularities at frequencies $\omega_j= \Delta/\hbar + j \omega_D$, which are associated with the generation of $j$ hard plasmons. Thus for $|\omega-\omega_j| \ll \omega_D$ we get 
\begin{equation}
W(\omega) \sim s_j(\omega-\omega_j) \left|\omega-\omega_j\right|^{\eta_j},
\end{equation}
where
\begin{equation} \label{eq:eta}
\eta_j = \frac{\delta p^2}{\pi m \hbar \omega'_0} + \frac{j}{2}-1,
\end{equation}
and $s_j(\omega)= \alpha_j+ \beta_j\theta(\omega)$ is a step function between two values which depend on $j$.

The existence of higher frequency singularities, $j>0$, Eq.~(\ref{eq:eta}), is a result of the harmonic approximation, which is valid as $r_{s}\rightarrow \infty$. At finite values of $r_{s}$ these singularities are smeared by anharmonic interactions between plasmons. The power of the higher frequency singularities, $\eta_j$, increases with $j$. Therefore only the first few of them are significant. An example of this behavior in the case where $\omega_q= \omega_D |\sin(q/2)|$ and the value of $\delta p$ is chosen such that $\eta_0=-0.4$ is depicted in Fig.~3. The  power law singularity at the absorption threshold ($j=0$) is not broadened by anharmonic interactions. Its exponent is $\eta_0  \approx \frac{\delta p^2}{n^2 \sqrt{r_s}}  -1$.

Depending on the momentum shift between the subbands, the photovoltage diverges or vanishes near the threshold. Finally we remark that the mass difference between the subbands, and the electron-phonon interaction do not change our qualitative picture. The mass difference will slightly increase the singularity powers $\eta_j$ due to the sudden change of the effective interaction strength between the excited electron and the rest of electrons, which leads to a contribution to $\eta_0$ associated with the orthogonality catastrophe.

Let us turn now to the case $r_s \ll 1$ where interactions are weak and electrons can move freely along the system. In this limit the threshold frequency approaches the single particle value $\omega_c^{sp}$ and usually $\delta p \ll p_F$ where $p_F$ is the Fermi momentum of the occupied band. The photoexcitation induces a direct transition which does not change the electron momentum. Therefore the M\"ossbauer mechanism described above does not apply. However, if the energy relaxation time of an excited electron in the upper subband is shorter than the momentum relaxation time of the electrons in the bottom band, then each photoexcitation results in an effective momentum transfer of $- p_F$ to the electrons in the lower band. Thus the photovoltage is given by  Eq.~(\ref{voltage}) in which $\delta p$ should be replaced by $p_F$.
The absorption rate has a power law singularity exponent at the absorption threshold, $W(\omega) \sim(\omega-\omega_c^{sp})^{\eta_0}$ with the exponent  $\eta_0 \sim r_s$.  The above threshold singularities shown in Fig.~3 are expected to smear out.

\section{V. Summary}

In this paper we have shown that the inclusion of interactions into the  CPGE problem  in CN changes the frequency dependence of the photovoltage $V(\omega)$ dramatically. Instead of a jump at the threshold frequency, the interaction produce a singularity with a power exponent which depends on  the relation between the momentum shift between subbands and the strength of the interaction between neighboring electrons, similar to the classical M\"{o}ssbauer effect.  We also show that $V(\omega)$ exhibits additional singularities at higher frequencies. These results provide a new experimental probe  by which the nature of electronic correlations in carbon nanotubes can be examined.

\begin{acknowledgments}
We thank E.~L.~Ivchenko, T.~Giamarchi, A.~Kamenev, and K.~Matveev for useful discussions. This research has been supported by the United States-Israel
Binational Science Foundation (BSF) grant No. 2008278 and by the U.S. DOE grant DE-FG02-07ER46452.
\end{acknowledgments}

\end{document}